\documentclass[twoside,twocolumn,9pt]{article}
\usepackage{extsizes}
\usepackage[super,sort&compress,comma]{natbib} 
\usepackage[version=3]{mhchem}
\usepackage[left=1.5cm, right=1.5cm, top=1.785cm, bottom=2.0cm]{geometry}
\usepackage{balance}
\usepackage{times,mathptmx}
\usepackage{sectsty}
\usepackage{graphicx} 
\usepackage{lastpage}
\usepackage[format=plain,justification=raggedright,singlelinecheck=false,font={stretch=1.125,small,sf},labelfont=bf,labelsep=space]{caption}
\usepackage{float}
\usepackage{fancyhdr}
\usepackage{fnpos}
\usepackage[english]{babel}
\usepackage{array}
\usepackage{droidsans}
\usepackage{charter}
\usepackage[T1]{fontenc}
\usepackage[usenames,dvipsnames]{xcolor}
\usepackage{setspace}
\usepackage[compact]{titlesec}
\usepackage{textcomp}
\usepackage{multirow}
\usepackage{todonotes}
\usepackage{soul}

\definecolor{cream}{RGB}{222,217,201}

\begin{document}

\pagestyle{fancy}
\thispagestyle{plain}
\fancypagestyle{plain}{

\fancyhead[C]{\includegraphics[width=18.5cm]{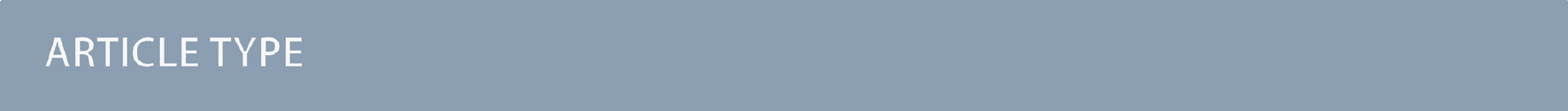}}
\fancyhead[L]{\hspace{0cm}\vspace{1.5cm}\includegraphics[height=30pt]{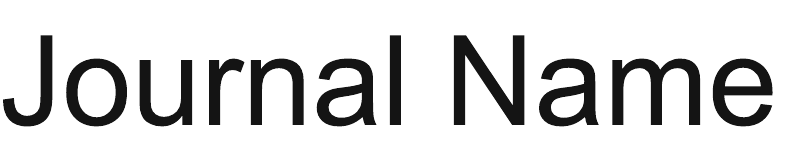}}
\fancyhead[R]{\hspace{0cm}\vspace{1.7cm}\includegraphics[height=55pt]{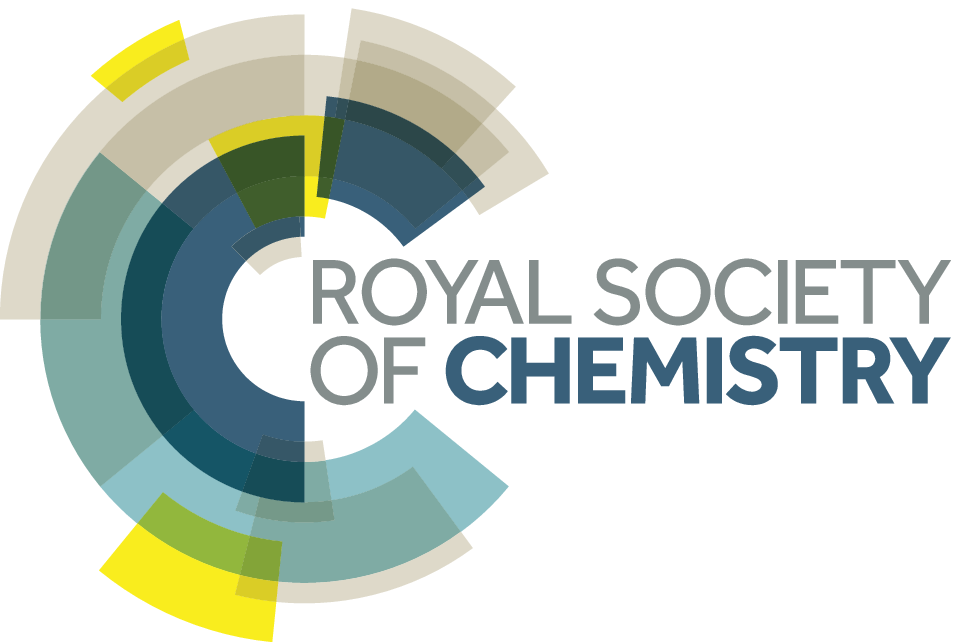}}
\renewcommand{\headrulewidth}{0pt}
}

\makeFNbottom
\makeatletter
\renewcommand\LARGE{\@setfontsize\LARGE{15pt}{17}}
\renewcommand\Large{\@setfontsize\Large{12pt}{14}}
\renewcommand\large{\@setfontsize\large{10pt}{12}}
\renewcommand\footnotesize{\@setfontsize\footnotesize{7pt}{10}}
\makeatother

\renewcommand{\thefootnote}{\fnsymbol{footnote}}
\renewcommand\footnoterule{\vspace*{1pt}%
\color{cream}\hrule width 3.5in height 0.4pt \color{black}\vspace*{5pt}} 
\setcounter{secnumdepth}{5}

\makeatletter 
\renewcommand\@biblabel[1]{#1}            
\renewcommand\@makefntext[1]%
{\noindent\makebox[0pt][r]{\@thefnmark\,}#1}
\makeatother 
\renewcommand{\figurename}{\small{Fig.}~}
\sectionfont{\sffamily\Large}
\subsectionfont{\normalsize}
\subsubsectionfont{\bf}
\setstretch{1.125} 
\setlength{\skip\footins}{0.8cm}
\setlength{\footnotesep}{0.25cm}
\setlength{\jot}{10pt}
\titlespacing*{\section}{0pt}{4pt}{4pt}
\titlespacing*{\subsection}{0pt}{15pt}{1pt}

\fancyfoot{}
\fancyfoot[LO,RE]{\vspace{-7.1pt}\includegraphics[height=9pt]{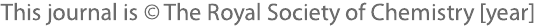}}
\fancyfoot[CO]{\vspace{-7.1pt}\hspace{13.2cm}\includegraphics{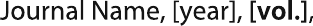}}
\fancyfoot[CE]{\vspace{-7.2pt}\hspace{-14.2cm}\includegraphics{RF}}
\fancyfoot[RO]{\footnotesize{\sffamily{1--\pageref{LastPage} ~\textbar  \hspace{2pt}\thepage}}}
\fancyfoot[LE]{\footnotesize{\sffamily{\thepage~\textbar\hspace{3.45cm} 1--\pageref{LastPage}}}}
\fancyhead{}
\renewcommand{\headrulewidth}{0pt} 
\renewcommand{\footrulewidth}{0pt}
\setlength{\arrayrulewidth}{1pt}
\setlength{\columnsep}{6.5mm}
\setlength\bibsep{1pt}

\makeatletter 
\newlength{\figrulesep} 
\setlength{\figrulesep}{0.5\textfloatsep} 

\newcommand{\topfigrule}{\vspace*{-1pt}%
\noindent{\color{cream}\rule[-\figrulesep]{\columnwidth}{1.5pt}} }

\newcommand{\botfigrule}{\vspace*{-2pt}%
\noindent{\color{cream}\rule[\figrulesep]{\columnwidth}{1.5pt}} }

\newcommand{\dblfigrule}{\vspace*{-1pt}%
\noindent{\color{cream}\rule[-\figrulesep]{\textwidth}{1.5pt}} }

\makeatother

\twocolumn[
  \begin{@twocolumnfalse}
\vspace{3cm}
\sffamily
\begin{tabular}{m{4.5cm} p{13.5cm} }

\includegraphics{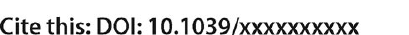} & \noindent\LARGE{\textbf{High-resolution structure of coexisting nanoscopic and microscopic lipid domains$^\dag$}} \\
\vspace{0.3cm} & \vspace{0.3cm} \\

 & \noindent\large{Michal Beli\v{c}ka,$^{\ast}$\textit{$^{a,b}$} Anna Weitzer,\textit{$^{a,b}$} and Georg Pabst$^{\ast}$\textit{$^{a,b}$}} \\ 

\includegraphics{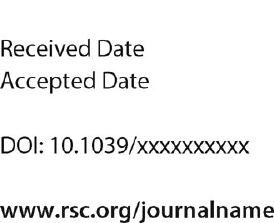} & \noindent\normalsize{We studied coexisting micro- and nanoscopic liquid-ordered/liquid-disordered domains in fully hydrated multilamellar vesicles using small-angle X-ray scattering. Large domains exhibited long-range out-of-plane positional correlations of like domains, consistent with previous reports. In contrast, such correlations were absent in nanoscopic domains. Advancing a global analysis of the \textit{in situ} data allowed us to gain deep insight into structural and elastic properties of the coexisting domains, including the partitioning of cholesterol in each domain. In agreement with a previous report we found that the thickness mismatch between ordered and disordered domains decreased for nanoscopic domains. At the same time we found the lipid packing mismatch to be decreased for nano-domains,  mainly by liquid-disordered domains becoming more densely packed when decreasing their size.} \\ 

\end{tabular}

 \end{@twocolumnfalse} \vspace{0.6cm}

  ]

\renewcommand*\rmdefault{bch}\normalfont\upshape
\rmfamily
\section*{}
\vspace{-1cm}


\footnotetext{$ ^* $ \textit{Corresponding authors. E-mail: michal.belicka@uni-graz.at, georg.pabst@uni-graz.at}}
\footnotetext{\textit{$^{a}$~University of Graz, Institute of Molecular Biosciences, Biophysics Division, NAWI Graz, Humboldtstr. 50/III, A-8010 Graz, Austria.}}
\footnotetext{\textit{$^{b}$~BioTechMed-Graz, A-8010 Graz, Austria.}}

\footnotetext{\dag~Electronic Supplementary Information (ESI) available: see DOI: 10.1039/b000000x/}



\section{Introduction}
\label{sec:intro}

Biological membranes are well-known to define and control intra- and intercellular environments.
One of the most intriguing features of biomembranes is their lateral heterogeneity,\cite{gaus_visualizing_2003,baumgart_large-scale_2007,lingwood_plasma_2008} called domains or rafts,\cite{jacobson_lipid_2007} which have been postulated to be involved in a wide range of physiological processes.\cite{simons_lipid_2000,lingwood_lipid_2010} Efforts for providing direct experimental evidence for the existence of membrane rafts have not been without significant controversy,\cite{kraft_plasma_2013,sevcsik_gpi-anchored_2015} which is often attributed to their small, nanoscopic size and/or short life-times.\cite{pike_rafts_2006}

In turn, membrane domains are well-established in complex lipid-only mixtures of low- and high-melting lipids and cholesterol (Chol),\cite{marsh_cholesterol-induced_2009} or other sterols which are able to condense sa\-tu\-rated hydrocarbons.\cite{mcconnell_condensed_2003,pandit_complexation_2004} Lipid-only domains of so-called liquid ordered ($ L_O $) and liquid disordered ($ L_D $) phases have been stu\-died by a wide range of experimental techniques, including fluorescence microscopy~(see, \textit{e.g.}, ref. \citenum{bagatolli_two_2000}), F\"{o}rster resonance energy transfer (FRET)~(see \textit{e.g.}, ref. \citenum{heberle_comparison_2010}), neutron diffraction,\cite{armstrong_observation_2013} small-angle X-ray and neutron scattering (SAXS, SANS),\cite{marquardt_scattered_2015} or nuclear magnetic resonance~(see \textit{e.g.}, ref. \citenum{veatch_liquid_2004}). $ L_O $ domains are con\-si\-de\-red as archetypes of rafts and their structural and elastic properties are of particular interest for understanding selective protein partitioning.\cite{gandhavadi_structure_2002,frewein_lateral_2016}

One of the interesting features of cholesterol-containing raft-like lipid mixtures is the ability to control domain size by lipid composition.\cite{feigenson_phase_2009} For example ternary mixtures with diunsaturated or highly branched lipids as the low-melting component display micron-sized domains, which can be readily observed under a microscope using fluorescent lipid labels (see, \textit{e.g.}, ref. \citenum{veatch_closed-loop_2006}). Exchanging the diunsaturated lipids to monounsaturated lipids, like palmitoyloleoylphosphocholine (POPC), instead reduces the size of these domains to a few nanometers.\cite{heberle_comparison_2010,heberle_hybrid_2013,ionova_phase_2012, konyakhina_phase_2013} Alternatively such systems have been proposed to resemble a microemulsion.\cite{schick_membrane_2012} Deciphering intrinsic structural properties of nanoscopic $L_O/L_D$ domains is highly challenging.
Most recently Nickels and coworkers\cite{nickels_mechanical_2015} were able to describe the thickness and bending rigidity of coexisting $L_O$ and $L_D$ domains in the nanoscopic regime in large unilamellar vesicles (LUVs) applying neutron scattering in combination with contrast variation.

Here we set out to show that detailed structural and elastic information of nanoscopic lipid domains in multilamellar vesicles (MLVs) can be retrieved from a global SAXS data analysis. Besides domain thickness and thickness of the hydrocarbon chains layer, our analysis is capable of retrieving the maximum bending fluctuation amplitudes of domains, as well as the packing density of lipids and the partitioning of cholesterol in $L_O/L_D$ domains.

To achieve this goal, we followed the approach of Heftberger \textit{et al.},\cite{heftberger_situ_2015} who detailed the analysis of coexisting micron-sized domains in MLVs by combining a modified Caill\'e theory description of the structure factor,\cite{zhang_theory_1994} with a scattering length density profile (SDP) model\cite{kucerka_lipid_2008} for the form factor. Our modeling led us also to reconsider domain stacking in the microscopic and nanoscopic regimes allowing for contributions of partially anticorrelated domains and overlapping domain leaflets. Additional modifications of our previous modeling included an intrinsic definition of the area per lipid (area per unit cell) and a generic parameterization of cholesterol distribution within the coexisting domains, which allows to extract cholesterol partitioning with high fidelity. Our SDP model for cholesterol differs from previous descriptions\cite{chen_partition_2007,heftberger_situ_2015} and is applicable for any other additive molecule(s) displaying affinity to either $ L_D $ or $ L_O $ phase. The new model yielded improved results for high-resolution SAXS data on micron-sized domains and gave first insights on structural properties of $ L_D $ and $ L_O $ domains in the nanoscopic regime. Most interestingly we found that the previously reported decrease of $ L_O $/$ L_D $ thickness mismatch for nanoscopic domains\cite{heberle_hybrid_2013} goes in hand with a decrease of the lipid packing mismatch.

\section{Theory}
\label{sec:theory}

Following the approach of Heftberger \textit{et al.}\cite{heftberger_global_2014,heftberger_situ_2015} we adopted the global analysis  model of Pabst \textit{et al.}\cite{pabst_structural_2000,pabst_structural_2003} for coexisting domains. For completeness and clarity of arguments we revisit some of the most important arguments leading to an improved modeling that also captures nanoscopic domains.

\begin{figure}[t]
	\centering
	\includegraphics[width=8.3cm]{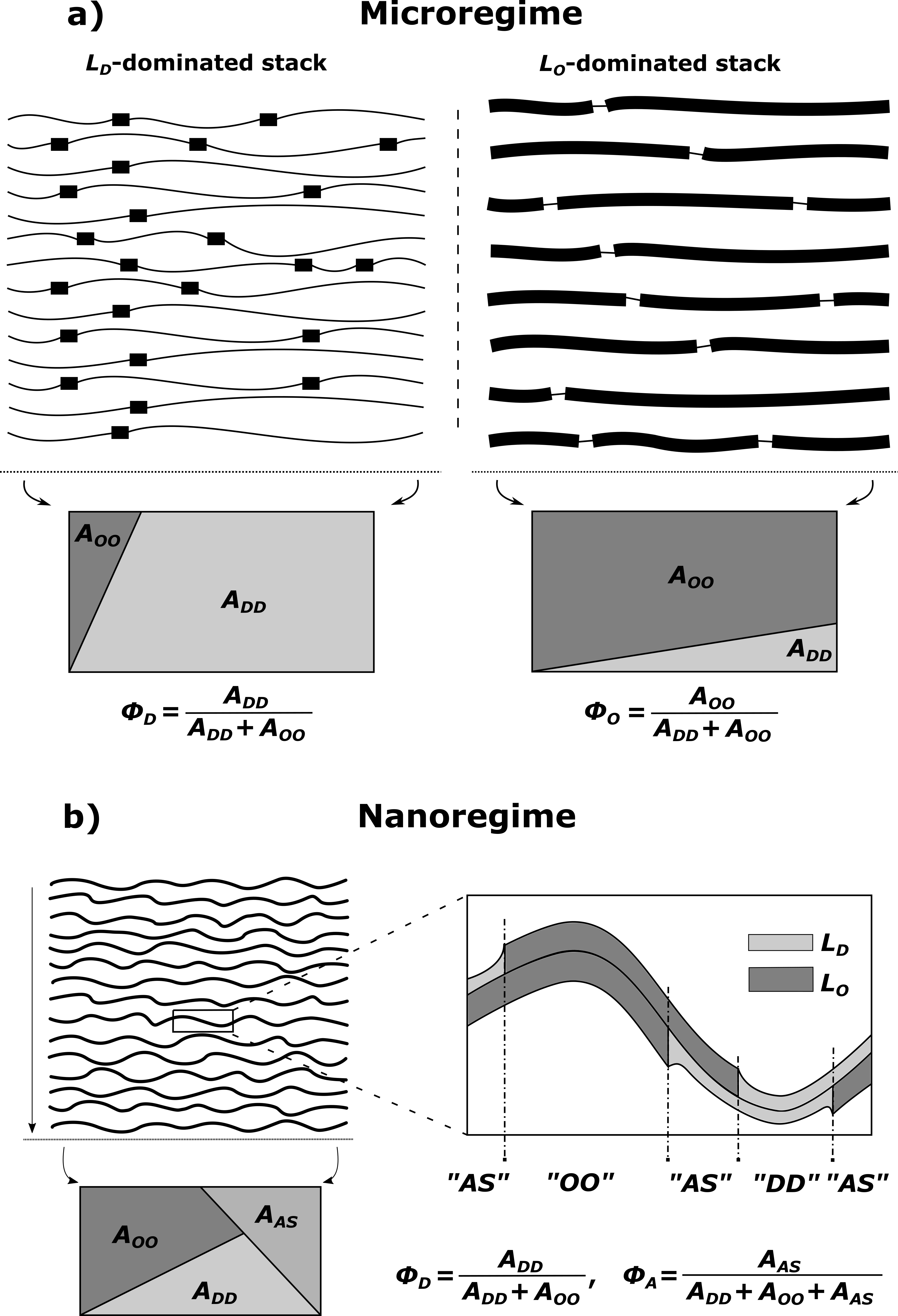}
	\caption{Schematic of domain alignment in the microscopic (A) and the nanoscopic regimes (B). In the microscopic regime each stack is dominated by either $L_D$ or $L_O$ domains, considering also contributions from unlike domains. This is taken into account by purity parameters $\Phi_D$, $\Phi_O$, which are based on the area fractions of ordered $A_{OO}$ and disordered $A_{DD}$ domains. In the nanoscopic regime an additional area fraction of leaflet anticorrelated domains $A_{AS}$ need to be taken into account in the purity parameter $\Phi_D$ (see text for details).}\label{fig:domains}
\end{figure}

MLVs displaying coexisting domains were consi\-de\-red as stacks of weakly bound bilayers averaged over all possible spatial orientations. Micron-sized domains were shown to exhibit long-range alignment of like domains in the stacking direction (see \textit{e.g.}, refs. \citenum{karmakar_phase_2006,chen_partition_2007,tayebi_long-range_2012}), giving rise to two distinct families of lamellar Bragg lattices, which we consider to be dominated by either $ L_O $ or $ L_D$ domains. That is, unlike previous assumptions, we presently also consider contributions from positionally anticorrelated domains, i.e. $ L_O $ in $ L_D $ and \textit{vice versa} (Fig.~\ref{fig:domains}a). The total scattering of such systems is
\begin{equation}\label{eq:Iinit}
I_{tot}(q)\approx l_D \cdot I_D(q) + l_O \cdot I_O(q),
\end{equation}
where $ l_D $ and $ l_O $ are the volume fractions occupied by $ L_D $ phase and $ L_O $ phase `dominated' stacks ($ l_D + l_O = 1 $), respectively, $ I_{D,O}(q) $ are the scattering intensities of the respective stacks, and $ q $ is the modulus of transferred momentum.

For nanoscopic domains only a single lamellar lattice is observed (see Results section). That is $ I_{tot}(q) $ contains contributions from $ L_O $ and $ L_D $ domains, but unlike MLVs with micron-sized domains, the interbilayer separation between $ L_O $ and $ L_D $ domains is identical (Fig.~\ref{fig:domains}b).

Contributions to the scattered intensity in MLVs are typically split into the structure factor $ S(q) $, describing the dynamic positional correlations between the layers and the form factor $ F(q) $, which accounts for the transbilayer/domain structure. Thus, for each stack of correlated bilayers in either the micro- or nanoscopic regime
\begin{equation}\label{eq:Istack}
\begin{split}
I^{corr}(q)=\frac{1}{q^2}\Big[S(q)&\cdot \lVert\langle F(q) \rangle\rVert^2 +\\
&+ \langle N\rangle \cdot (\langle\lVert F(q) \rVert^2\rangle - \lVert\langle F(q) \rangle\rVert^2)\Big],
\end{split}
\end{equation}
where $ \langle N\rangle $ is the average number of positionally correlated bilayers in stacks within irradiated volume.

Additional scattering contributions from positionally uncorrelated bilayers/domains, originating \textit{e.g.} from defects due to packing mismatches, need to be considered.\cite{pabst_structural_2003} Taking these into account the full scattering intensity of each stack becomes
\begin{equation}\label{eq:Idiff}
\begin{split}
I_{s}(q)=\frac{1}{q^2}\Big[(1-f_{diff})I^{corr}(q) + f_{diff}\lVert\langle{}F(q)\rangle\rVert^2 \Big],
\end{split}
\end{equation}
where $ f_{diff} $ gives the fraction due to diffuse scattering. Naturally, $ f_{diff} $ comprises also of the scattering contributions from unilamellar vesicles, that may be present in the sample.

\subsection{Structure factor}
\label{subsec:sf}

Positional correlations of fluid multibilayers are known to be affected by membrane undulations, which can be described by the modified Caill\'{e} theory (MCT).\cite{zhang_theory_1994,zhang_small-angle_1996} Within this theory the structure factor for each stack is
\begin{equation}\label{eq:smct}
\begin{split}
S_{MCT}(q;N) = N&+2\sum_{k=1}^{N-1}\Big\{(N-k)\cos(kqd)\times\\
&\times\exp\Big(-\big(d/2\pi\big)^2q^2\eta\big[\gamma+\ln(\pi k)\big]\Big)\Big\}, 
\end{split}
\end{equation}
where $ N $ is the number of spatially correlated bilayers per scattering domain, $ \gamma $ is Euler's constant and $ \eta $ is the Caill\'e pa\-ra\-me\-ter, which is a measure of bilayer fluctuations, that depends on the bending rigidity and interbilayer interactions.\cite{gennes_physics_1993} From $ \eta $ we can directly calculate the fluctuations in bilayer separation $\Delta_{fl}^2=\eta d^2/\pi^2$.~\cite{petrache_structure_1998}
The polydispersity of $ N $ can be described by different probability distributions.\cite{zhang_theory_1994,zhang_small-angle_1996,petrache_structure_1998,fruhwirth_structure_2004} Here we chose an exponential probability distribution
\begin{equation}
f(N; N_{mean})=\frac{1}{N_{mean}}\exp\Big(-\frac{N}{N_{mean}}\Big),
\end{equation}
which is a special case of the well-known Schulz-Flory distribution,\cite{aragon_theory_1976,bartlett_neutron_1992,pencer_method_2006} where $ N_{mean} $ is the sample average of $ N $. We also tested a regular Schulz-Flory distribution, but the exponential distribution yielded better fits for all presently studied systems.

\subsection{Form factor}
\label{subsec:ff}

To model the form factor, we have to consider two different scenarios (Fig.~\ref{fig:domains}): (i) leaflet-correlated domains ($ L_O $/$ L_O $, $ L_D $/$ L_D $) and (ii) leaflet-anticorrelated domains ($ L_O $/$ L_D $, $ L_D $/$ L_O $). Micron-sized domains are well known to exhibit transbilayer coupling of like domains.\cite{blosser_transbilayer_2015} Thus, only leaflet-correlated domains need to be considered. In the nanoscopic regime both types of domains may occur as suggested by a recent simulation study.\cite{fowler_roles_2016} In general, the form factor is given by the Fourier transformation of the \textit{contrast scattering length density profile} $ \Delta\rho(z) $ $ (=\rho(z) - \rho_s) $ along the bilayer normal, $ z $,
\begin{equation}\label{eq:formfac}
F(q)=\int_{-d/2}^{d/2}\Delta\rho(z)e^{-iqz}dz,
\end{equation}
where $ \rho(z) $ is the transbilayer electron density and $ \rho_s $ is the electron density of the solvent. Note that leaflet-correlated domains represent a centrosymmetric crystalline system, while leaflet-anticorrelated domains are asymmetric.

In order to easily differentiate between symmetric and asymmetric systems, we describe the transdomain structure for each leaflet separately. For modeling the internal leaflet structure we applied the scattering length density profile (SDP) description developed by Ku\v{c}erka~\textit{et al.},\cite{kucerka_lipid_2008,kucerka_fluid_2011} which allows to analyze jointly X-ray and neutron scattering data. In the framework of the SDP model, each lipid molecule is parsed into quasi-molecular fragments. For the presently studied lipids these components are the choline methyl groups (ChoMet), phosphate+CH$ _2 $CH$ _2 $N (PCN), glycerol+carbonyls (GC), hydrocarbon methylene groups (HC) and finally the terminal chain methyls (CH3).

The crystallographic unit cell is therefore realized by a single phospholipid molecule, which is in the present experimental case given by an averaged DSPC/DOPC, or DSPC/POPC molecule. Cholesterol is considered as an additive, which is able to distribute into $ L_D $ and $ L_O $ phases as discussed in subsection~\ref{subsec:moldist}. The base area of this unit cell is an important structural parameter -- the \textit{interfacial area per lipid molecule A}. In the case of lipid mixtures with cholesterol $A$ also contains a fraction of cholesterol and is therefore referred to as the \textit{area per unit cell}.~\cite{kucerka_influence_2007} Naturally this applies to any admixture molecule. Note that this definition differs from the partial lipid areas reported in other studies.~\cite{edholm_areas_2005,hodzic_differential_2008}
Each individual component of the lipid molecule is represented in the SDP model by its \textit{lateral unit fraction} $ a_i(z) $ -- a function describing the fractional contribution of a given component to $A$, requiring that $\sum_{i} a_i(z) = 1$ due to ideal volume filling. 

For each leaflet the contribution of the hydrocarbon chain fraction is described by a single \textit{plateau}-function
\begin{equation}
a_{HC}(z)=0.5\Bigg[\textrm{erf}\bigg(\frac{z}{\sqrt{2\sigma_{C}^2}}\bigg)-\textrm{erf}\bigg(\frac{z-D_C}{\sqrt{2\sigma_{C}^2}}\bigg)\Bigg],
\end{equation}
where $ D_C $ and $ \sigma_{C} $ are the position and ``roughness'' of the hydrocarbon chain region interface, respectively and $\textrm{erf}(x) $ refers to the error function. In our SDP model the unit HC region volume $ V_{HC} $ contains also a fraction of lipophilic additives,
\begin{equation}
\label{eq:admixture}
V_{C}=V_{chains}+\sum_i r_i\cdot V_{adm,i},
\end{equation}
where $ V_{chains} $ is the total volume of the phospholipid hydrocarbon chains (i.e. a mixture of low-melting and high-melting lipids, excluding carbonyl groups), $ V_{adm,i} $ is the volume and $ r_i $ is the re\-la\-tive molar ratio of $ i^{th} $ lipophilic additive, here given by cholesterol. The calculation of $ r_i $ for $ L_O $ and $ L_D $ domains is detailed in subsection \ref{subsec:moldist}. Because of the assumed conservation of volumes, $ D_C $ is coupled to $ A $ through
\begin{equation}\label{eq:a_z}
A=\frac{V_{C}}{D_C}.
\end{equation}
Thus $ A $ and $ D_C $ are not interdependent parameters. In the present work we fitted $ A $ and derived $ D_C $ through Eq.~(\ref{eq:a_z}).


All other quasi-molecular fragments of the unit lipid molecule, \textit{i.e.}, the polar headgroup components and the terminal methyl group, are given by Gaussian probability densities\cite{kucerka_lipid_2008} with an average \textit{volume distribution}
\begin{equation}
v_i(z) = \frac{V_i}{\sqrt{2\pi\sigma_i^2}}\exp\bigg(-\frac{(z-z_i)^2}{2\sigma_i^2}\bigg),
\end{equation}
where $ V_i $ is the molecular volume of the $ i^{th} $ component ($i =$ PCN, CG, \ldots), $ z_i $ is the center of mass, and $ \sigma_i $ is the Gaussian standard deviation. Note that $ V_i $ needs to be scaled by the relative molar ratio $ r_i $ according to the concentration of cholesterol (or any admixture molecule in general), $ V_i = r_i\cdot V_{i,single} $, where $ V_{i,single} $ are the volumes reported for the pure lipid components.
The cor\-res\-pon\-ding $ a_i(z) $'s are then given by
\begin{equation}\label{eq:ai}
a_i(z)=\frac{v_i (z)}{A}.
\end{equation}

The final transbilayer SDP is constructed by a step-by-step replacement of the homogeneous water ($ \rho_w $)/hydrocarbon core ($ \rho_{HC} $) background in both leaflets by properly scaled individual components (using Eq.~\ref{eq:ai})
\begin{eqnarray}\label{eq:rhoprof}
\begin{aligned}
\rho(z)=\rho_w+\sum_{\textrm{\textit{m} in water}}a_m(z)(\rho_m-\rho_w)\quad +\\
+\sum_{\textrm{\textit{k} in HC}}a_k(z)(\rho_k-\rho_{HC}),
\end{aligned}
\end{eqnarray}
where $\rho_{m,k} =n_{m,k}^e/V_{m,k}$ are the electron densities of a given quasi-molecular fragment with $n_{m,k}^e$ number of total electrons, and $\rho_{HC}=\rho_{CH_2}$. The first sum in Eq.~(\ref{eq:rhoprof}) runs through all components within the water environment (including the hydrocarbon chains core) and the second one through all hydrophobic components (\textit{e.g.}, terminal methyl groups, cholesterol).

Because of minute differences in hydrocarbon volumes of presently studied phospholipids we were not able to deconvolute their contributions. Therefore, molecular averages of phospholipids in each domain were calculated according to reported compositional phase diagrams.\cite{heberle_comparison_2010,konyakhina_phase_2013} Since lipid headgroup volumes do not change significantly with temperature or phase,\cite{marsh_handbook_2013} the vo\-lu\-mes and widths of the PC headgroup components were assumed to be the same in both phases. Their values were adopted from the works of Klauda~\textit{et al.}\cite{klauda_simulation-based_2006} and Ku\v{c}erka~\textit{et al.}.\cite{kucerka_fluid_2011} The choline methyls carry insufficient electron density contrast to be detected by X-rays. We therefore fixed their widths and coupled their relative positions to the CG group $ z_{CG} = D_C + 0.9\,\textrm{\AA} $.\cite{kucerka_fluid_2011}

The hydrocarbon chain volumes were calculated using data reported in refs. \citenum{marsh_handbook_2013} and \citenum{uhrikova_component_2007}, assuming that
\begin{equation}
\frac{V_{CH}(L_O)}{V_{CH_2}(L_O)}=\frac{V_{CH}(L_D)}{V_{CH_2}(L_D)},
\end{equation}
and $ V_{CH_2}(L_O) $ equals that reported for the gel phase (see Tab.~S2 in \textit{Supplementary material}). Further, we assumed $ V_{CH_3} = 2 V_{CH_2} $ in the $ L_D $-phase,\cite{marsh_handbook_2013} while $ V_{CH_3} $ was an adjustable parameter in $ L_O $ domains.

In order to describe the scattering of microscopic or nanoscopic domains in MLVs, we now need to consider contributions from `impurities'. In the case of micron-sized domains these originate from unlike domains such as $ L_O $ in $ L_D $ dominated stacks and \textit{vice versa} (Fig.~\ref{fig:domains}a). The average area fraction of each domain in each stack is described by the \textit{purity} $ \Phi_i $ ($ \Phi_D, \Phi_O $). For definition, see Fig.~\ref{fig:domains}. The amplitudes of the averaged form factors and their averaged squared amplitudes in Eqs.~\ref{eq:Istack} and \ref{eq:Idiff} then become
\begin{eqnarray}\label{eq:mixffmicro1}
\begin{aligned}
\lVert\langle F^D(q) \rangle\rVert &=\lVert \Phi_DF_D(q) + (1-\Phi_D)F_O(q)\rVert, \\
\lVert\langle F^O(q) \rangle\rVert &=\lVert \Phi_OF_O(q) + (1-\Phi_O)F_D(q)\rVert,
\end{aligned}
\end{eqnarray}
and
\begin{eqnarray}\label{eq:mixffmicro2}
\begin{aligned}
\langle\lVert F^D(q)\rVert^2\rangle &=\Phi_D\lVert F_D(q)\lVert^2 + (1-\Phi_D)\lVert F_O(q)\rVert^2, \\
\langle\lVert F^O(q)\rVert^2\rangle &=\Phi_O\lVert F_O(q)\lVert^2 + (1-\Phi_O)\lVert F_D(q)\rVert^2.
\end{aligned}
\end{eqnarray}

In the case of nanoscopic domains, we do not find long-range alignment of like domains (see section~\ref{subsec:resnano}), but need to consider contributions from leaflet-anticorrelated domains (Fig.~\ref{fig:domains}b). Thus, we have scattering from symmetric ($ L_O $/$ L_O $, $ L_D $/$ L_D $) and asymmetric domains ($ L_O $/$ L_D $, $ L_D $/$ L_O $). Since we have no measure for orientation of the asymmetric domains, we assume that half of the asymmetric domains are $ L_O $/$ L_D $ and the other half $ L_D $/$ L_O $. Defining the area fraction of asymmetric domains, $ \Phi_A $, as illustrated in Fig.~\ref{fig:domains}b, we find for the form factor amplitudes of nanoscopic domains   
\begin{eqnarray}\label{eq:mixffnano1}
\begin{aligned}
\lVert\langle F^{nn}(q)\rangle\rVert &= \lVert \Phi_A F_A(q) + \\
&+(1-\Phi_A)\big(\Phi_D F_D(q) + (1-\Phi_D)F_O\big)\rVert
\end{aligned}
\end{eqnarray}
and
\begin{eqnarray}
\begin{aligned}
\langle\lVert F^{nn}(q)\rVert^2\rangle &= \Phi_A\lVert F_A(q)\rVert^2 + \\
&+ (1-\Phi_A)\big(\Phi_D \lVert F_D(q)\Vert^2 + (1-\Phi_D)\lVert F_O(q)\rVert^2\big),
\end{aligned}
\end{eqnarray}
where $ F_D(q) $, and $ F_O(q) $ are the form factors of symmetric $ L_D $ or $ L_O $ domains, respectively, and $ F_A(q) $ is the form factor of asymmetric domains.

\subsection{Partitioning of cholesterol}
\label{subsec:moldist}

The effect of cholesterol (and any other added lipophilic molecule) on the scattering curve depends not only on the overall surface coverage of a given domain, but even more crucially on its molar ratio/fraction within $ L_O $ or $ L_D $. Thus, the cholesterol content of a given domain can be determined from scattering experiments. Recently, Ma \textit{et~al.}\cite{ma_cholesterol_2016} estimated cholesterol partitioning in $ L_D / L_O $ domains by fitting Gaussian-like functions to electron density profiles, assuming that observed profile changes originate only from the added electron densities of cholesterol.

Here we introduce an alternative approach, which does not rely on the above mentioned assumption. The relative molar ratios of cholesterol in $ L_D $ and $ L_O $ domains ($ r_D, r_O $) can be decoupled from the overall molar ratio of cholesterol in the lipid mixture $ r_{tot} = n_{chol} / n_{lip} $, where $ n_{chol} $ and $ n_{lip} $ are the total molar numbers of cholesterol and phospholipid per sample, by considering
\begin{equation}
r_{tot}=\frac{n_{lip,D}\cdot r_D + n_{lip,O}\cdot r_O}{n_{lip,D} + n_{lip,O}},
\end{equation}
with
\begin{eqnarray}
n_{lip,D}=N_D\frac{l_D\Phi_D}{A_D}A_{ir}+N_O\frac{(1-l_D)(1-\Phi_O)}{A_D}A_{ir}, \\
n_{lip,O}=N_D\frac{l_D(1-\Phi_D)}{A_O}A_{ir}+N_O\frac{(1-l_D)\Phi_O}{A_O}A_{ir}
\end{eqnarray}
in the microscopic regime (see Eq.~(\ref{eq:Iinit}) for the definition of $ l_D $) and
\begin{eqnarray}
n_{lip,D}=N \frac{0.5\cdot{}\Phi_A+(1-\Phi_A)\Phi_D}{A_D}A_{ir}, \\
n_{lip,O}=N \frac{0.5\cdot{}\Phi_A+(1-\Phi_A)(1-\Phi_D)}{A_O}A_{ir}
\end{eqnarray}
in the nanoscopic regime. Here, $ A_{ir} $ is the average surface area of a single sheet within a given stack of bilayers, $ A_D $/$ A_O $ are the areas of the unit cells in $ L_D $/$ L_O $ domains, and $ N_D $/$ N_O $ are the average numbers of positionally correlated bilayers in each stack.

Inserting Eq.~(20) into Eq.~(19) and Eq.~(22) into Eq.~(21) allows to remove $ A_{ir} $ and one obtains after some arithmetic
\begin{equation}
r_O=r_{tot} + (r_{tot} - r_D)\cdot \Gamma,
\end{equation}
where
\begin{equation}\label{eq:rcholmicro}
\Gamma = \frac{N_Dl_D\Phi_D+N_O(1-l_D)(1-\Phi_O)}{N_Dl_D(1-\Phi_D)+N_O(1-l_D)\Phi_O}\cdot\frac{A_O}{A_D},
\end{equation}
for the microscopic regime and
\begin{equation}\label{eq:rcholnano}
\Gamma =\frac{0.5\Phi_A+(1-\Phi_A)\Phi_D}{0.5\Phi_A+(1-\Phi_A)(1-\Phi_D)}\cdot\frac{A_O}{A_D},
\end{equation}
for the nanoscopic regime, which were used to evaluate Eq.~(\ref{eq:admixture}). Specifically, we fitted $ r_{D} $ and calculated $ r_{O} $ using the equations above. Heftberger \textit{et~al.}\cite{heftberger_situ_2015} used a similar model, but fitted $r_{D}$ and $r_{O}$ independently. 

The molecular volume of cholesterol in $ L_D $ domains $ V_{Chol}(L_D) $ = 628\,\AA$ ^3 $ was supplied from volumetric measurements of DOPC/Chol mixtures.\cite{gallova_partial_2015} In turn, $ V_{Chol} $ is not known for $ L_O $ domains. We therefore decided to vary $ V_{Chol}(L_O) $ between its known extremes in binary mixtures.\cite{marsh_handbook_2013} In most cases the final value of $ V_{Chol}(L_O) $ ended up between 600\,\AA$ ^3 $ and 650\,\AA$ ^3 $.

In each domain cholesterol is parsed into a head and acyl tail group.\cite{heftberger_structure_2015} We further assumed that the electron density of the acyl tail group $\rho_{Chol}^t = \rho_{HC}$, from which we calculate the corresponding volume $ V_{Chol}^t = n_e^t/\rho_{Chol}^t $, with $ n_e^t $ being the number of electrons of the cholesterol tail. The cholesterol head group volume is simply derived from $V_{Chol} = V_{Chol}^h+V_{Chol}^t$. The cholesterol head group is represented by a Gaussian-like $ a_{Chol}(z) $ (\ref{eq:ai}). All parameters supplied from previous measurements are summarized in Tab.~S2. The overall number of adjustable parameters varied between 18 in the microscopic regime and 13 in the nanoscopic one. The application of a global search algorithm was therefore necessary in order to avoid getting stuck in a local minimum during optimization. Details of the implementation of the model and applied optimization algorithm are given in section 4 of the supplementary material.

For comparison to other structural data we define the Luzzati thickness as a measure for the thickness of the domains as\cite{nagle_structure_2000}
\begin{equation}\label{eq:db}
D_B=2\frac{V_{L}}{A},
\end{equation}
where $V_L$ is the total lipid volume. Finally, the head-to-head distance of a given domain is calculated from the cross-bilayer distance between the PCN groups
\begin{equation}\label{eq:dhh}
D_{HH}=2z_{PCN}.
\end{equation}

\section{Materials and methods}
\label{sec:materials}

\subsection{Chemicals}
\label{subsec:chems}

POPC, DOPC and DSPC were purchased from Avanti Polar Lipids (Alabaster, AL, USA) and cholesterol from Sigma-Aldrich (Vienna, Austria) as dry powders. All lipids were used without any further purification. Organic solvents of spectral purity were obtained from Lactan (Graz, Austria) and Milli-Q water (18\,M$ \Omega $ cm at 25\,\textdegree{C}) was freshly prepared before use.

\subsection{Sample preparation}
\label{subsec:prep}

Dispersions of fully hydrated MLVs were prepared by rapid solvent exchange as detailed by Rieder \textit{et al.}.\cite{rieder_optimizing_2015} Briefly, first stock solutions were prepared by dissolving weighted amounts of lipid in methanol/chloroform (1/9) solvent. Lipid concentration was determined  to $<$ 1\% by inorganic phosphate assay.\cite{kingsley_synthesis_1979} Lipid mixtures of POPC\,:\,DSPC\,:\,Chol = 0.39\,:\,0.39\,:\,0.22 and DOPC\,:\,DSPC\,:\,Chol = 0.46\,:\,0.3\,:\,0.24 were obtained by mixing appropriate amounts of the stock solutions. Then 300 $\mu$l of each solution was transferred into a test tube containing 600~$ \mu $l Milli-Q water using a gastight Hamilton syringe. The test tube was mounted onto the RSE apparatus and heated under a constant stream of argon to 70\,\textdegree{C} (above the boiling points of methanol and chloroform). This allowed the fast evaporation of the organic solvent without the need of negative pressure. Each mixture was evaporated for 12 minutes. During the evaporation samples were constantly vortex-mixed at 1000 rpm to prevent sedimentation of methanol-chloroform solution droplets as well as to increase their evaporation rate. 
The final lipid concentration in each sample was at least 30\,mg/ml.

\subsection{Measurements}
\label{subsec:measure}

The SAXS scattering curves were obtained either at the ESRF BM29 BioSAXS beamline~\cite{pernot_upgraded_2013} (Grenoble, France) or at the P12 SAXS beamline~\cite{blanchet_versatile_2015} at DESY (Hamburg, Germany). At BM29 samples were measured at an X-ray wavelength of $ \lambda $ = 0.99\,\AA\/ using a sample-detector-distance (SDD) of 2.869\,m, whereas experiments at the P12 beamline were performed at $ \lambda $ = 0.6\,\AA\/ and SDD = 3.1\,m. Scattered intensities were recorded using a Pilatus 1M (BM29), and a Pilatus 2M (P12) detector (Dectris, Baden, Switzerland).

At both beamlines samples were transferred prior to measurement into multi-well plates and equilibrated for 10 minutes in a temperature-controlled block. An automated sample robot de\-li\-ve\-red 20-35\,$\mu$l of the lipid sample into a preheated glass capillary. For each sample 20 frames were recorded with an exposure time of 0.095 s at P12 and 10 frames, each with 1~s exposure, at BM29.  Water background was measured before and after each sample. To avoid introductions of artifacts by radiation damage, data collected in subsequent frames were compared and rejected in case of statistically significant deviations. Background subtraction was performed by using the ATSAS software suite.\cite{petoukhov_new_2012}

\section{Results and Discussion}
\label{sec:results}

Prior to applying our model to coexisting domains, we tested its capabilities on well-studied single component bilayers of pure DOPC. Our best fit agreed well with the scattering data obtained at 20\,\textdegree{C} over the whole $q$-range. The resulting structural parameters are detailed in Tab. S1 and the corresponding fit is plotted in Fig.~S1.

Comparing to literature values we focus in particular on the la\-te\-ral area per lipid. From our analysis we obtained $ A = 64.3 $\,\AA$ ^2 $, which agrees favorably with $ A = 66.9 $\,\AA$ ^2 $~\cite{kucerka_structure_2006} and $ A = 67.6 $\,\AA$ ^2 $,\cite{heftberger_global_2014} both obtained at 30\,\textdegree{C}, considering the lateral expansion of bilayer with temperature~\cite{kucerka_fluid_2011}. We therefore conclude that our model successfully captures also high-resolution structural information of single component bilayers. For further structural results, see \textit{Supplementary material}.

\subsection{Microscopic domains}
\label{subsec:resmicro}

\begin{figure}[t]
	\centering
	\includegraphics[width=8.3cm]{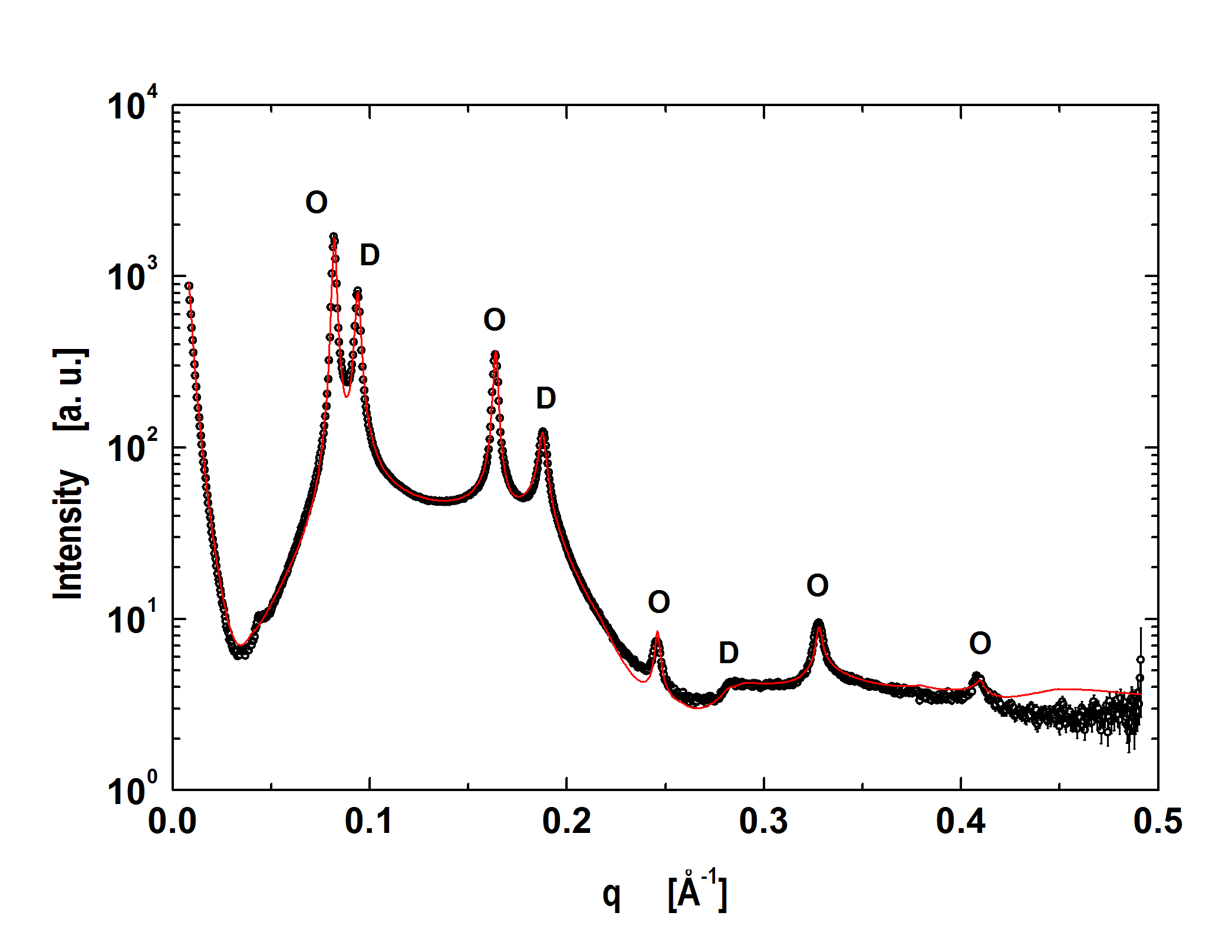}
	\caption{Global analysis (red line) of microscopically phase separated DOPC/DSPC/Chol multibilayers. Peaks corresponding to $ L_D $ dominated stacks are indicated by `D's' and `O's' correspond to the lamellar lattice of $ L_O $ dominated stacks.}\label{fig:MICROfit}
\end{figure}

The next step was to evaluate the model for micron-sized domains, which have been shown to exhibit long-range out-of-plane domain alignment in multibilayers and which have been analyzed in terms of a global SAXS data analysis before.~\cite{heftberger_situ_2015,kollmitzer_bending_2015} Thus our system of choice was the above detailed mixture of DOPC/DSPC/Chol (0.46/0.3/0.24).~\cite{heberle_bilayer_2013} The corresponding data shown in Fig.\,\ref{fig:MICROfit} clearly show the presence of two co\-existing lamellar lattices, where the higher $d$-spacing phase (76.7\,\AA) is ascribed to $ L_O $-dominated stacks, implying that the lattice of $ d = 66.9 $\,\AA{} corresponds the $ L_D $-dominated stacks. These values are in excellent agreement with our previous report.\cite{heftberger_situ_2015} 

\begin{figure}[!t]
	\centering
	\includegraphics[width=8.3cm]{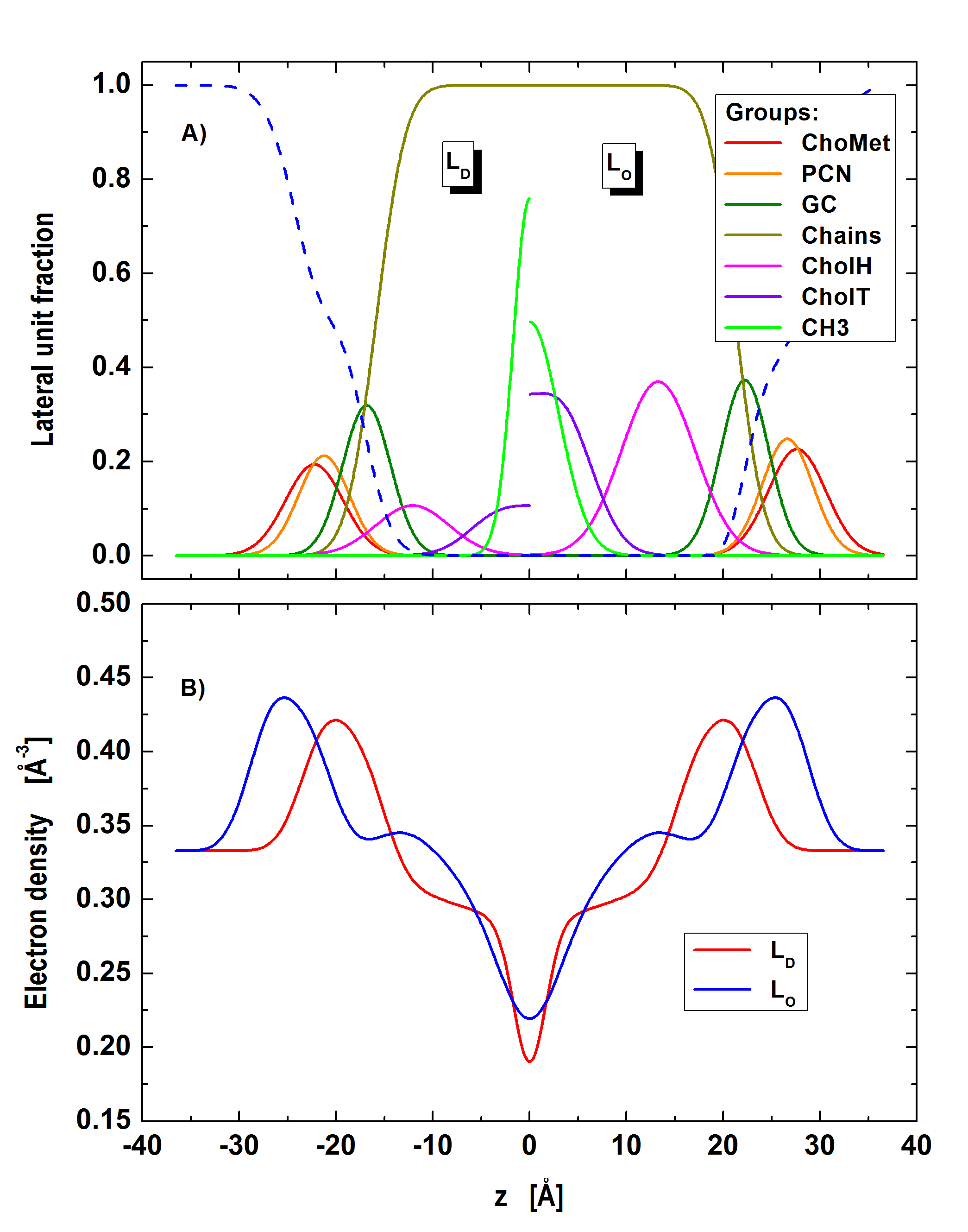}
	\caption{Distribution of component groups (A) and corresponding electron density profiles (B) of DOPC/DSPC/Chol $ L_D $ and $ L_O $ domains in the microscopic regime.}
	\label{fig:MICROprof}
\end{figure}

Our model is able to fit the data well up to $q < 0.4$\,\AA$^{-1}$. Discrepancies for $ q > 0.4 $\,\AA$ ^{-1} $ are most probably caused by subtle structural features of the $ L_O $ polar headgroup region, which we currently do not capture. The resulting group distribution functions of $ L_D $ and $ L_O $ phases and the corresponding electron density profiles are presented in  Fig.\,\ref{fig:MICROprof}.  

Lattice parameters obtained from our detailed analysis (Tab.~\ref{tab:microstrf}) show that about two thirds of the sample ($ l_D = 0.675 $) were composed of $ L_D $ dominated stacks. Looking at the abundance of each phase we found that $ L_D $-stacks were somewhat `purer' in $ L_D $ domains ($ \Phi_D $ = 93.6\%) than $ L_O $ stacks in $ L_O $ domains ($ \Phi_O $ = 73.1\%).

The relative molar ratios of cholesterol were found to be $ r_O = 0.493 $ in $ L_O $ domains and $ r_D = 0.208 $ in $ L_D $ domains. These data compare well with $ r_D = 0.205 $ and $ r_O = 0.471 $ reported for the same tie-line from FRET studies by Heberle \textit{et al.},\cite{heberle_comparison_2010} lending strong support to the applied modeling. We emphasize that the obtained results for $ r_D $ and $ r_O $ were stable without the need to apply constraints.

Bending fluctuations were significantly lower in $L_O$ domains, \textit{i.e.}~by a factor of $ \sim 1.75 $, than in $ L_D $ domains, which is due to the increased content of DSPC and cholesterol.~\cite{heberle_bilayer_2013} Our $ \eta $ values were, however, systematically larger than reported previously,\cite{heftberger_situ_2015} what can be attributed to the presently higher experimental resolution and the improved model description. Interestingly, bending fluctuations of $ L_D $ domains, which are rich in DOPC, are significantly higher (factor: $ \sim 2.4 $) than those of pure DOPC MLVs (see Tab. S1.), despite the fact that they contain some cholesterol. This might be due to local bilayer thinning defects at the domain boundaries as suggested by Nickels \textit{et al.},~\cite{nickels_mechanical_2015} leading to increased domain fluctuation amplitudes.

\begin{table}[t]
	\centering
	\caption{MLVs lattices structure parameters in the microscopic regime. Parameter uncertainties are $<$ 2\%.}\label{tab:microstrf}
	\begin{tabular}[]{|ccc|}
		\hline
		\multirow{2}{*}{\textbf{Parameter}} & \multicolumn{2}{c|}{\textbf{Bilayer lattice}} \\
		& $ L_D $-dominated & $ L_O $-dominated \\
		\hline
		$ l_i $ & 0.67  & 0.33 \\
		$ \Phi_i $ & 0.94  & 0.73 \\
		$ d $  [\AA] & 66.9 & 76.6 \\
		$ \eta \times 10^{-2} $ & 13.4 & 3.3 \\
		$ \Delta_{fl} $  [\AA] & 7.8  & 4.5 \\
		$ N_{bil} $ & 10.5 & 10.9 \\
		\hline
	\end{tabular}
\end{table}

\begin{table}[!t]
	\begin{center}
		\caption{Structural parameters of $ L_D $ and $ L_O $ domains in the microscopic regime. Parameter uncertainties are $<$ 2\%.}\label{tab:microff}
		\begin{tabular}[]{|ccc|}
			\hline
			\multirow{2}{*}{\textbf{Parameter}}& \multicolumn{2}{c|}{\textbf{Microdomains}} \\
			& $ L_D $ domains & $ L_O $ domains \\
			\hline
			$ A $  [\AA$^2$] & 69.5  & 57.4  \\
            $ A^m $  [\AA$^2$] & 57.5 & 38.4 \\
			$ D_C $  [\AA] & 16.0  & 21.2 \\
			$ D_B $  [\AA] & 41.5  & 53.9 \\
			$ D_{HH} $ [\AA] & 42.6 & 53.4 \\
			$ z_{GC} $  [\AA] & 16.9 & 22.2 \\
			$ z_{PCN} $  [\AA] & 21.3  & 26.7 \\
			$ r_{Chol,i} $ & 0.208 & 0.493 \\
			$ z_{Chol} $  [\AA] & 12.1 & 13.3 \\
			$ \sigma_{CH_3} $  [\AA] & 1.94 & 2.63 \\
			\hline
		\end{tabular}
	\end{center}
\end{table}

Turning to domain structure (Tab.~\ref{tab:microff}) and comparing to our previous report,\cite{heftberger_situ_2015} the new $ A $ va\-lues are somewhat higher for both domains ($ L_D $: $ A = 69.5 $\,\AA$ ^2 $ \textit{vs.} $ 60.3 $\,\AA$ ^2 $ and $ L_O $: $ A = 57.4 $\,\AA$ ^2 $ \textit{vs.} $ 43.1 $\,\AA$ ^2 $), especially for $ L_O $. We believe, that the discrepancies are mainly caused by different ways of the calculation of $ A $. Here, we fitted $ A $ assuming volume conservation of the hydrocarbon core, considering also contributions from cholesterol (\ref{eq:a_z}), while Heftberger \textit{et al.}\cite{heftberger_situ_2015} calculated $ A $ by finding the Luzzati thickness from the water distribution function. 
Further, the improved quality of data and model fits may also explain the the observed differences. The increase of $ A $ in $L_D$ domains, with respect to pure DOPC (Tab. S1) is due to the fraction of cholesterol contributing to the unit cell, consistent with a previous report on binary mixtures with cholesterol.~\cite{kucerka_influence_2007} 
Defining the \textit{average lateral area per molecule} as $A^m = A/n_m$, where $n_m$ is the number of molecules per unit cell (including both PCs and Chol), we find a tighter packing of the molecules in both domains ($A^m_D = 57.5 $\,\AA$^2$ \textit{vs.} $ A^m_O = 38.4$\,\AA$^2$) as compared to pure DOPC, which naturally agrees with the commonly accepted lateral membrane condensation property of cholesterol. The lower $ A $-value of $ L_O $ domains as compared to $L_D$ domains is consistent with an enrichment in cholesterol and DSPC.

Because $ D_C $ is coupled to $ A $ (Eq.~\ref{eq:a_z}), our hydrocabon core thickness values are smaller than reported previously ($ \sim $ 2 \AA~for $ L_D $ and $ \sim $ 5 \AA~for $ L_O $).\cite{heftberger_situ_2015} However, when comparing the Luzzati thicknesses, these differences are minimal, in particular for $ L_O $ domains (present: $ D_B (L_D)  = 41.5 $~\AA, $ D_B(L_O)  = 53.9 $~\AA; Heftberger \textit{et al.}:\cite{heftberger_situ_2015} $ D_B (L_D) = 39.2 $~\AA, $ D_B(L_O) = 49.2 $~\AA) and almost agree within experimental resolution.

Finally, the average positions of the cholesterol headgroup were almost identical in both phases, $ z_{Chol}(L_D) = 12.1 $\,\AA~and $ z_{Chol}(L_O) = 13.3 $\,\AA, which is in good agreement with our previous report using a simpler parsing description for cholesterol.\cite{heftberger_situ_2015}

\subsection{Nanoscopic domains}
\label{subsec:resnano}

\begin{figure}[t]
	\centering
	\includegraphics[width=8.3cm]{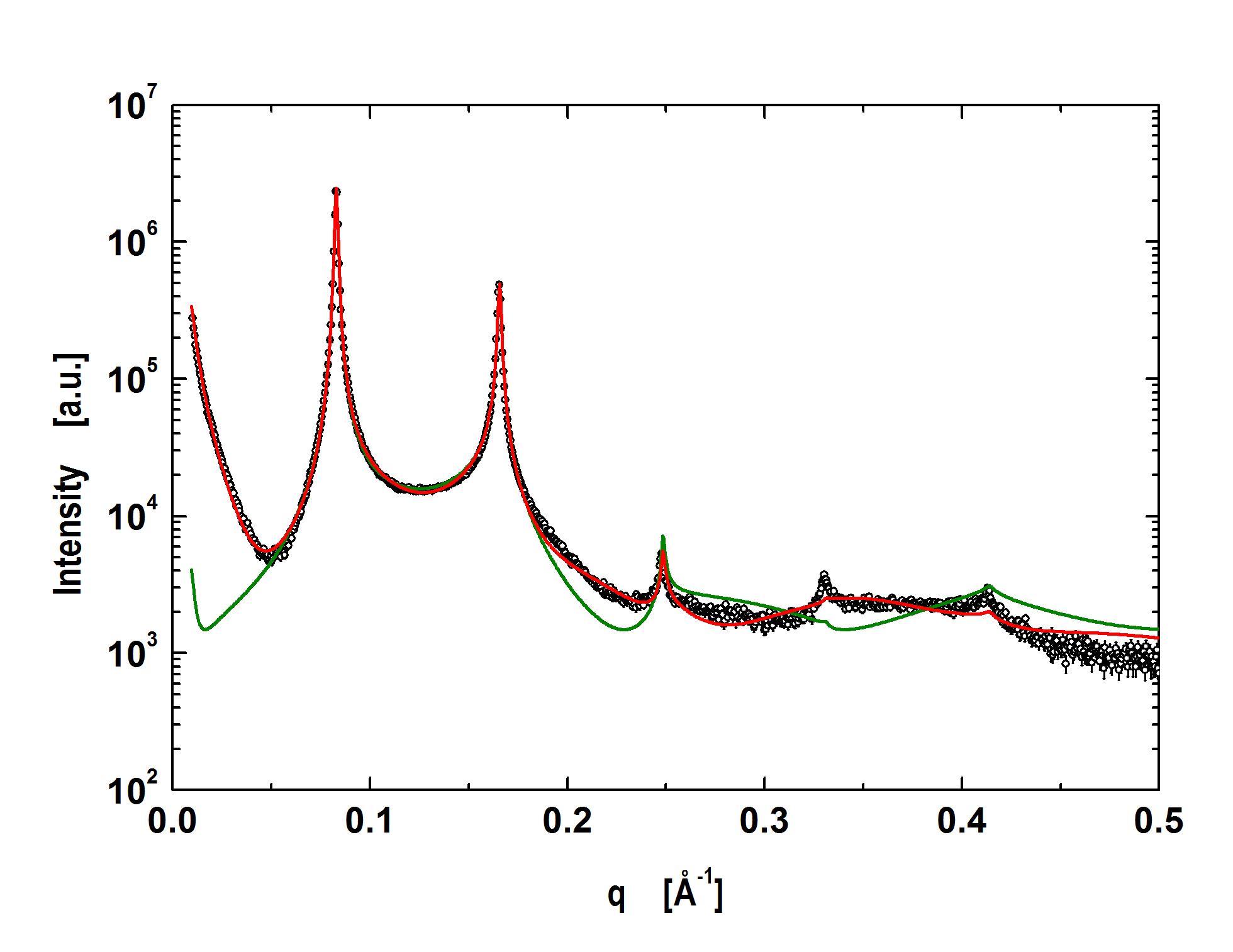}
	\caption{Global analysis (red line) of coexisting nanoscopic domains in POPC/DSPC/Chol multibilayers. The green line shows the best fit assuming  homogeneous lipid mixing.}\label{fig:NANOfit}
\end{figure}

Nanoscopic domains in POPC/DSPC/Chol (0.39/0.39/0.22), did not exhibit long-range out-of-plane positional correlations of like domains, but displayed only a single lamellar lattice (Fig.\,\ref{fig:NANOfit}) with a $ d $-value of $ 75.9 $\,\AA, \textit{i.e.} very close to the $ d $ of micron-sized $ L_O $ domains (Tab.~\ref{tab:microstrf}). To exclude the possibility of the randomly-mixed bilayers, we also carried out a corresponding single phase fit. However, even the best single phase fit was not able to follow the scattering curve (Fig.\,\ref{fig:NANOfit}). This encouraged us to utilize the model of MLVs containing nanoscopic domains.

\begin{figure}[!t]
	\centering
	\includegraphics[width=8.3cm]{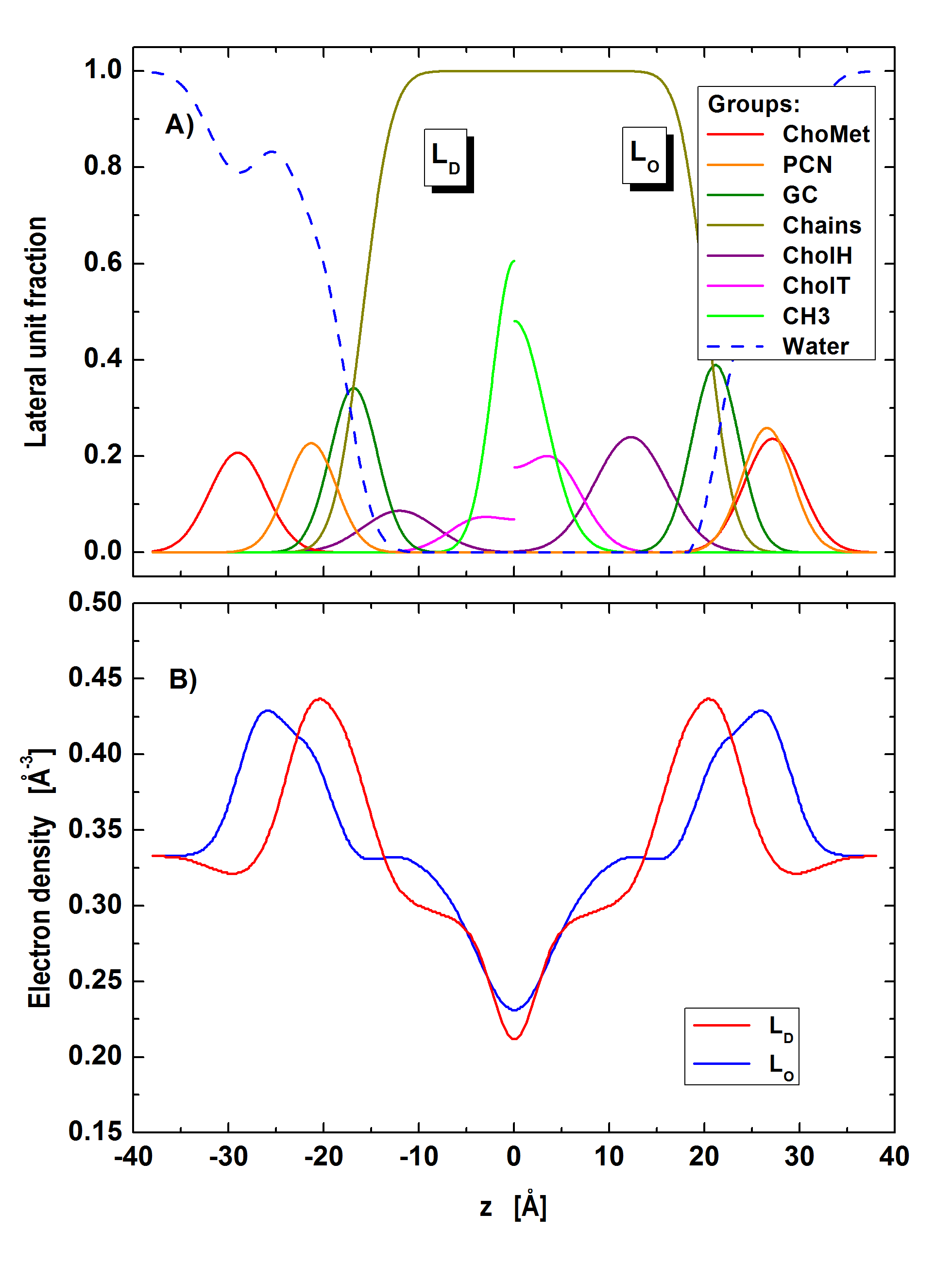}
	\caption{Distribution of component groups (A) and corresponding electron density profiles (B) of POPC/DSPC/Chol $ L_D $ and $ L_O $ domains in the nanoscopic regime.}
	\label{fig:NANOprof}
\end{figure}

During fitting with the nanoscopic domain model we found instabilities of the optimization primarily with cholesterol related parameters (position, concentration). This is most likely due to the small contrast  between nanoscopic $L_O$ and $L_D$ domains for X-rays. Further exploitation of this issue requires a joint analysis of differently contrasted SANS data with SAXS data of nanoscopic domains. Such studies are currently being planed in our laboratory. Note that the presently described analysis is capable of performing this analysis without any further modification.
To analyze current SAXS data we tested different sets of different structural and volumetric constraints for cholesterol. The best and stable fit was obtained upon fixing (i) the relative cholesterol molar ratios ($ r_D = 0.14, r_O = 0.34 $) according to reported tie-line endpoints of the same mixture,~\cite{heberle_bilayer_2013} (ii) the relative positions of cholesterol to GC groups, $ \Delta z_{Chol,i} = z_{GC,i} - z_{Chol,i} $, to the values obtained in the microscopic regime ($ \Delta z_{Chol,D} = 4.8 $\,\AA{}, $ \Delta z_{Chol,O} = 8.9 $\,\AA{}), and (iii) the cholesterol molecular volumes ($ V_{Chol,D} = V_{Chol,O} = 630 $\,\AA$^3$) in agreement with.~\cite{heberle_bilayer_2013}  The correspon\-ding fit is shown in Fig.~\ref{fig:NANOfit}, the  resulting domain structure is shown in Fig.~\ref{fig:NANOprof} and corresponding model parameters are summarized in Tabs.\,\ref{tab:nanostrf} and \ref{tab:nanoff}.

Regarding the lattice parameters (Tab.~\ref{tab:nanostrf}) we found that bending fluctuations, $\eta$, and their amp\-li\-tudes, $ \Delta_{fl}$, are close to those in micron-sized $ L_O $ domains. This is consistent with $ L_D $ domains being less abundant than $L_O$ domains, as obtained from our analysis  $ \Phi_D = 32\% $. This value is also in agreement with the reported compositional phase diagram.~\cite{heberle_comparison_2010} Interestingly, the found fraction of leaflet anti-correlated domains $ \Phi_A\sim $ 1\% is small. Despite this small value $\Phi_A$ has a significant effect on the scattered intensity. That is, setting $ \Phi_A = 0 $ does not produce a satisfactory fit of the SAXS pattern.

Regarding domain structure, we did also observe changes compared to micron-sized domains (Tab.~\ref{tab:nanoff}, for corresponding electron densities and component distribution functions, see Fig.\,\ref{fig:NANOprof}). For example the area per lipid of $ L_D $ domains reduced by about 7\%, while $ A_{O} $ decreased by about 2\%, only. That is, the mismatch in packing of lipids between $ L_D $ and $ L_O $, $\Delta A = |A_{D}-A_{O}|$  domains become much less expressed for nanoscopic domains ($\Delta A^{nano} \sim 8$ \AA$^2$, $\Delta A^{micro} \sim 12$ \AA$^2$). This may at least in part be due to the different composition of the presently studied nanoscopic $ L_D $ domains, which with POPC as the dominant species contain a lipid that has a smaller area per lipid than DOPC, enriched in `our' micron-sized $ L_D $ domains.\cite{kucerka_areas_2009,kucerka_fluid_2011}

Focusing on transdomain structure, we see the most significant changes for $ L_O $ domains with a decrease of the hydrocarbon chain length ($ D_C $) by $\sim 1.0 $\,\AA, and $ D_B $ by $\sim 1.8 $\,\AA, whereas thickness changes in $ L_D $ were significantly smaller. This yields in total a smaller thickness difference between $ L_O $ and $ L_D $ domains ($ \Delta D_B = D_B(L_O) - D_B(L_D) $) in the nanoscopic regime, when compared to the microscopic one ($\Delta D_B^{nano} \sim 10$ \AA, $\Delta D_B^{micro} \sim 12$ \AA), which is consistent with a previous report.\cite{heberle_hybrid_2013} Differences in absolute $ D_B $ values\cite{heberle_hybrid_2013,nickels_mechanical_2015} may be due to different model descriptions for the scattering length densities.

\begin{table}[t]
	\centering
	\caption{MLVs lattice structure parameters in the nanoscopic regime. Parameter uncertainties are $<$ 2\%.}\label{tab:nanostrf}
	\begin{tabular}[]{|cc|}
		\hline
		\textbf{Parameter} & \textbf{Bilayer lattice} \\
		\hline
		$ \Phi_D $ & 0.32 \\
		$ \Phi_A $ & 0.023 \\
		$ d $  [\AA] & 75.9 \\
		$ \eta \times 10^{-2} $ & 4.9 \\
		$ \Delta_{fl} $  [\AA] & 5.3 \\
		$ N_{bil} $ & 16.8 \\
		\hline
	\end{tabular}
\end{table}

\begin{table}[!t]
	\begin{center}
		\caption{Structural parameters of $ L_D $ and $ L_O $ domains in the nanoscopic regime. Parameter uncertainties are $<$ 2\%.}\label{tab:nanoff}
		\begin{tabular}[]{|ccc|}
			\hline
			\multirow{2}{*}{\textbf{Parameter}}& \multicolumn{2}{c|}{\textbf{Nanodomains}} \\
			& $ L_D $ domains & $ L_O $ domains \\
			\hline
			$ A $  [\AA$^2$] & 64.1 & 56.3  \\
            $ \hat{A} $  [\AA$^2$] & 56.2 & 42.0 \\
			$ D_C $  [\AA] & 15.9 & 20.2 \\
			$ D_B $  [\AA] & 42.2 & 52.1 \\
			$ D_{HH} $ [\AA] & 42.6 & 53.1 \\
			$ z_{GC} $  [\AA] & 16.9 & 21.1 \\
			$ z_{PCN} $  [\AA] & 21.3 & 26.5 \\
			$ r_{Chol,i}^* $ & 0.14 & 0.34 \\
			$ z_{Chol}^* $  [\AA] & 12.1 & 12.3 \\
			$ \sigma_{CH_3} $  [\AA] & 2.3 & 3.2 \\
			\hline
		\end{tabular}
	\end{center}
	
	$ ^{*)} $ Fixed values.
\end{table}

\section{Conclusions}
\label{sec:concl}

In the present paper we have advanced our previously reported SDP model\cite{heftberger_situ_2015} for \textit{in situ} studies of coexisting lipid domains in multibilayers. The presented model accounts for defects induced by positionally anticorrelated domains in micron-sized domains coexisting in MLVs. For nanoscopic domains no long-range alignment of like domains along the stacking direction was observed. However, modeling had to account for contributions from overlapping/asymmetric domains ($ L_D $/$ L_O $, $ L_O $/$ L_D $), which was inspired by a recent simulation report.\cite{fowler_roles_2016} Our modeling allowed us to capture a range of structural details, \textit{e.g.}~area per unit cell, domain thickness, \textit{etc.}, of coexisting microscopic and nanoscopic domains.

We found distinct differences in lipid packing densities and domain thicknesses in micron-sized domains in agreement with previous results.\cite{heftberger_situ_2015} These differences were found to be less expressed for nano-sized lipid domains, signifying a decrease of thickness mismatch between $ L_D $ and $ L_O $, in agreement with a previous neutron scattering study.~\cite{heberle_hybrid_2013} Here, we also found that the packing of lipids becomes more alike, in particular by laterally more condensed nanoscopic $L_D$ domains. For micron-sized domains our analysis allowed us to determine cholesterol partitioning in $ L_O $ and $ L_D $ phases which agreed well with published data from compositional phase diagrams.

We emphasize that deriving these structural details did neither involve the use of any labels, nor require measuring samples at tieline endpoints. 
Future studies will be extended to an SDP-based joint neutron and X-ray data analysis to fully exploit contrast variation, analogous to several pre\-vious reports on homogeneous lipid bilayers systems.\cite{pabst_applications_2010,marquardt_scattered_2015} This will allow us to test results obtained from current modeling and enable us to determine cholesterol content in nanoscopic domains.

\section*{Acknowledgements}

The authors are indebted to Frederick A. Heberle for valuable discussions and critical reading of the manuscript. The authors further thank Cl\'{e}ment Blanchet (EMBL Hamburg) and Adam Round (ESRF) for technical assistance. The research leading to these results has received funding from the Austrian Science Funds (FWF), project number I1304-B20 (to GP) and the European Community's Seventh Framework Programme (FP7/2007-2013) under BioStruct-X (grant agreement No. 6042.12). Measurements at BM29 were supported through the Austrian BAG proposal MX-1740.

\bibliography{mylibrary} 
\bibliographystyle{rsc} 

\end{document}